\documentclass[journal]{IEEEtran}
\usepackage{graphicx}
\DeclareGraphicsExtensions{.png,.pdf,.jpg}
\usepackage{cite}
\usepackage{orcidlink}

\begin{document}

\title{Satellite Underflight Utility for Thermal Sensor Harmonization}
%
%
%

\author{Sarah~ E. Kay\,\orcidlink{0000-0002-8281-8388} and 
        Brian~N. Wenny,~\IEEEmembership{Member,~IEEE,}
        
\thanks{Science Systems Applications Inc. (SSAI), Lanham, MD 20706.}
} 
\maketitle

\begin{abstract}
Underflight maneuvers provide a unique opportunity to harmonize calibration of on-orbit sensors. Due to their similar sensor technologies, their near-identical transmission profiles, orbital properties and platform operations, the underflight data of Landsat 8 and 9 instruments stand out as a qualifier to test proposed metrics, methods, and the extent over which to compare two independently calibrated sensors across their similar operating bandpasses. This study performed a pixel-to-pixel comparison of thermal imagery of TIRS and TIRS-2 (aboard Landsat 8 and 9, respectively) during their five-day underflight maneuver in November 2021, with the ultimate goal of identifying the key site/scene-selection criteria for a subset of images that are suitable for radiative calibration validation purposes. If a group of near-coincidentally observed images by two identical underflying sensors fail to show consistent Top-of-Atmosphere (TOA) Brightness Temperatures for the same exact geographical locations, then the scenes with those shared properties and/or observing conditions will prove unreliable for cross calibration validation of less-similar underflying sensor pairs.  This study demonstrates that near-coincidental images with Root Mean Square Deviations (RMSDs) of less than 5\% between their TOA radiances are optimum candidates for cross-validation of radiative calibration between two independently calibrated sensors. This criterion is shown to be reliable for coincidental acquisitions with a wide range of overlapping area, terrain type, land-to-water fraction and cloud coverage. Important considerations include any time gap between near-coincident acquisitions as well as the application of pixel quality masks. The analysis of the selected underflight scenes demonstrated an agreement between the TIRS on Landsat 8 and 9 to within 0.123 K and 0.066 K for the 10.9 $\mu m$ and 12.0 $\mu m$ bands respectively.
\end{abstract}


\IEEEpeerreviewmaketitle

\section{Introduction}
%
%
%
%
\IEEEPARstart{T}{he} successful launch of Landsat 9 on September 27, 2021 enabled the continuation of the 50+ year data record of Earth observation of the Landsat program. Cross-calibration and radiometric consistency between the Landsat sensors is a key element to the success and usefulness of the data archive to the worldwide science community. To that end, the launches of the Landsat 8 and 9 platforms were coordinated to perform an underflight maneuver of the current operational platform to allow near-simultaneous observations of the same Earth scene targets \cite{mar85,tei00,hel12,kil16,hol16,kai22,cho22,gro22}. Cross-calibrating and validating radiometric performance of the two platforms allows for a level of data harmonization to maintain consistency in the Landsat data products \cite{reu15}. 
There are various methods to coordinate maneuvers between platforms for simultaneous observations during which an orbiting platform either flies-by or maintains a co-moving pairing with another platform with orbits either above or underneath the said platform for a defined period. In the case of the Landsat 8 and 9 underflight event, the maneuver took place during the initial ascension of Landsat 9 to its operational WRS-2 orbit \cite{cho22}.  The ascension had Landsat 9’s orbital track crossing under that of Landsat 8 over a period of several days (November 11th to 16th, 2021). The near-coincidental observations of the same scenes largely eliminate the effects of rapid temporal changes such as the atmospheric influences on the sensor measurement and facilitates a more direct radiometric comparison between the two sensors.
Landsat 9 carries two instruments, the Operational Land Imager (OLI) and the Thermal Infrared Sensor (TIRS) which are essentially copies of the OLI and TIRS instruments on Landsat 8. The OLI has 9 visible and short-wave infrared bands (VSWIR) and TIRS has two thermal infrared bands, Landsat bands 10 and 11, nominally centered at 10.9 and 12 microns respectively. As the two OLI and TIRS instruments are nearly identical, the underflight dataset provides an opportunity to cross-compare the coincident measurements of two sensors where the potential complicating issues of instrument differences have been removed or minimized. Thus, this provides an ability to harmonize the data products at the earliest stage of the newly launched mission.  The initial analysis of the underflight data focused on the VSWIR measurements of the two OLI instruments and their geometric and radiometric performance \cite{cho22,gro22}. The TIRS sensors relied on observations of the traditional vicarious calibration ground targets (e.g. Lake Tahoe and Salton Sea) to harmonize the radiometric performance between the two TIRS \cite{mon14,bar14}. Coordinated field campaigns during the underflight compared the surface temperature products of Landsat 8 and Landsat 9 TIRS and found good agreement between the two well within 0.5 K for both water buoys and vicarious thermal measurements over several land surface types \cite{eon23}.
This work investigates a methodology for and analysis of a comparison of the top-of-atmosphere (TOA, defined as the outermost layer of the atmosphere that extends from about 600 to 10,000 km above the Earth) thermal measurements of the two TIRS instruments during the underflight period. Intuitively, one expects a comparison of two near-identical sensors viewing the same target coincidentally to agree very well with each other. In practice, subtle nuances in instrument performance, seemingly small differences in observation times or view angles and impacts of operational processing algorithms add challenges which are investigated. While performing an underflight at launch provides a point-in-time harmonization data point to maintain continuity of operational to newly launched sensors, the benefits of periodic underflights between pairs of sensors enables tracking sensor performance and identification of possible sensor issues. Given the number of current and planned thermal missions in different orbital tracks and altitudes, the value of utilizing this technique would benefit the science community. For example, the idea of purposely coordinating periodic underflights for the upcoming SBG-TIR, LSTM, and TRISHNA missions is under discussion.
The insights gained from the analysis presented here can inform data harmonization strategies for Earth observing missions. The introduced criteria are chosen such that they are adaptable to less-similar sensors pairs. We postulate that if the chosen criteria/metrics to verify calibration consistency for these two near-identical sensors failed to display radiometric consistency with high fidelity, then those metrics will inevitably exhibit even lower confidence and clarity for confirming the calibration consistency and/or flagging potential calibration issues.  Landsat platforms have had underflight data collection with published records of their use but despite the added value of the underflight data of Landsat 8 and 9 in particular, beyond satisfying the mission operation objectives \cite{kai22,cho22,gro22}, their full potential for investigative calibration validation purposes have not been utilized.  As calibration validation (Cal/Val) plans for the new generation of sophisticated broad-spectrum sensors are taking shape [e.g., see \cite{goo18,tur23,kud24, rob23, pou23, gra19, shr23, lib23}], comprehensive studies that highlight the value of the specialized datasets such as underflight data are more motivated than ever before. This study aims to fill the realized void in utilizing underflight data for on-orbit Cal/Val planning, site selection criteria and effective radiometric validation metrics.

\section{Data and Methodology}
\begin{figure}[!t]
\centering
\includegraphics[width=2.5in]{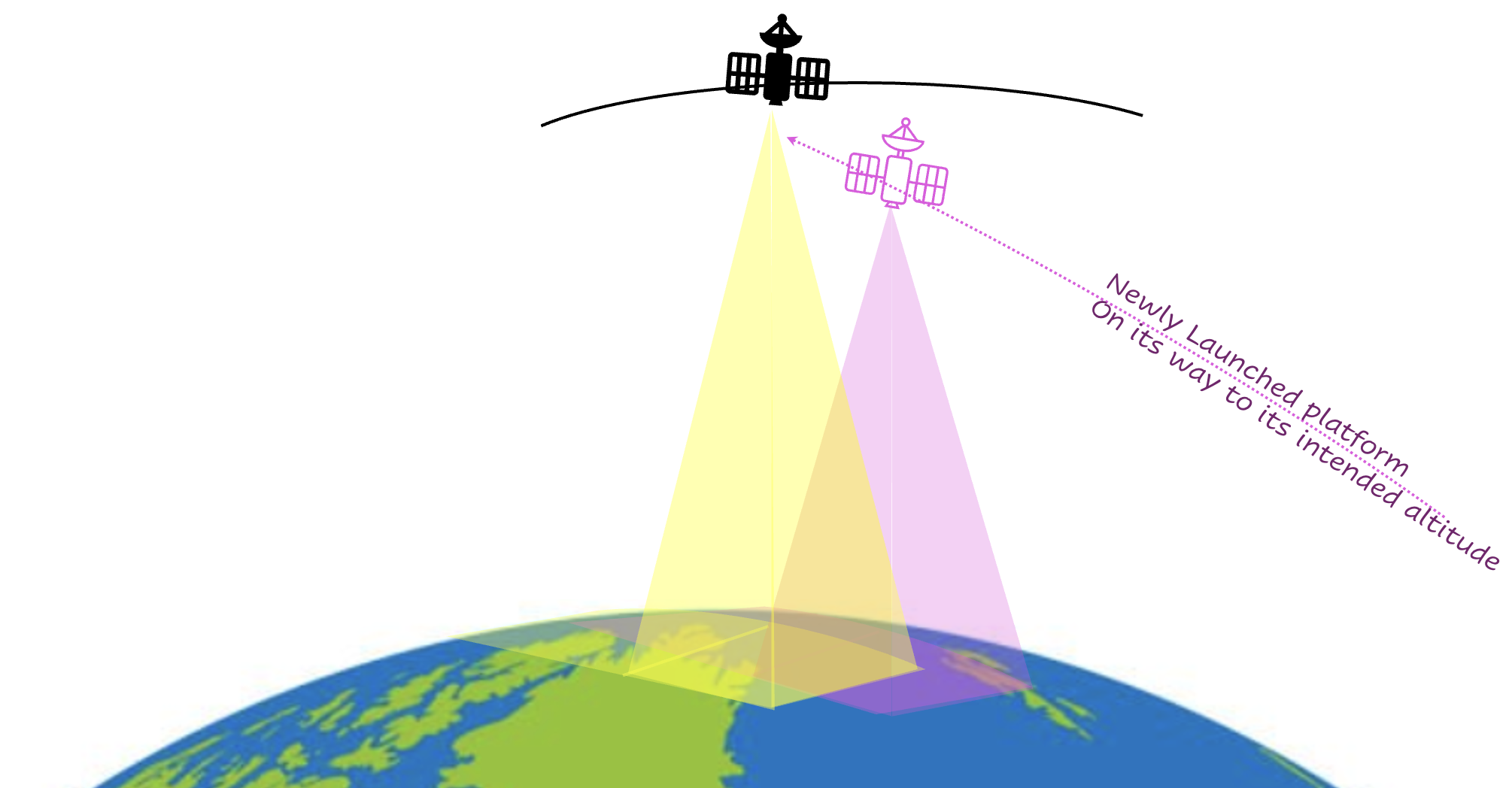}

\caption{Schematic illustration of the orbital tracks of Landsat 8 and Landsat 9 during the November 11-16, 2021, underflight operational period. As Landsat 9 was maneuvered to its final orbit, the ground track drifted east-to-west across the Landsat 8 ground track with the nearest simultaneity on November 14.
}
\label{fig1}
\end{figure}

Extensive planning prior to the launch of Landsat 9 in September 2021 resulted in a 5-day period of orbit overlap with Landsat 8. Operational constraints dictated the timing, location and duration of the underflight activity. The successful result of these planning efforts was a 5-day period with overlapped Earth-view scenes for Landsat 8 (L8) and Landsat 9 (L9) with time differences of less than 10 minutes and view angle differences of less than 20 degrees \cite{mar85}. Figure \ref{fig1} shows the general setup of the orbital tracks of L8 and L9 during the November 11-16, 2021, underflight operational period. As L9 was maneuvered to its final orbit, the ground track drifted east-to-west across the L8 ground track with the nearest simultaneity during November 14.  The orbit for maximum overlap was chosen to include specific vicarious ground targets such as Lake Tahoe. While the imagery between the two sensors is near simultaneous, the view angle of each TIRS can also be different for a given set of paired images depending upon the relative orbital positions between the two sensors. There can exist small observation time differences between paired images. This work restricts the selected data to those observed within 10 minutes of each other.
Since this study aims to create a pathway for more efficient use of the valuable underflight data, and advocate for its periodic use during a sensors’ operational lifetime, it is important to find the lowest level data product, that is readily available and yet sufficiently calibrated for a pixel-to-pixel comparison. The presented analyses in this study utilize the per-pixel TOA spectral radiance (TOARAD in $Watts/(m^{2}.srad.\mu m)$) and the corresponding brightness temperature (TBr in Kelvin) drawn from the Level 1 Terrain Precision (L1TP) product for both L8 and L9 TIRS instruments. L1TP images are quantized and calibrated digitized data products (i.e., arrays of digital counts, DN), representing the imaging data at each specific bandpass in the TIRS instruments (B10 and B11 centered at 10.9 and 12 microns respectively). The TIRS bands have a nominal spatial resolution of 100m, which is then resampled to generate the 30m L1TP products. The distortion effects in satellite imagery due to the Earth’s varying topography are corrected using digital elevation models (DEMs) to ensure spatial geometry of the images are as accurate as possible.
The per-pixel TOARAD values are computed at B10 and B11 in both TIRS instruments as:

\begin{equation}
TOARAD = M_{L} Q_{cal}+AL
\end{equation}
Where $M_L$ is the band-specific multiplicative rescaling factor, and $AL$ is band-specific additive rescaling factor from the metadata, and $Q_{cal}$ is the Quantized and calibrated standard product pixel values ($DN$).   The calculated $TOARAD$ values per pixel are converted to the $TOA$ brightness temperature using the thermal constants that are delivered in the associated Metadata ($MTL$) file with each image:

\begin{equation}
T_{Br} = \frac{K_2}{ln(\frac{K_1}{TOARAD}+1)}
\end{equation}
Where $K_1$ and $K_2$  are band-specific thermal conversion constants from the provided metadata per image \cite{eros22}.

Assuming that two overlapping pixel-arrays are sections of the same simultaneously observed earth region, high similarity between the two arrays of TOA radiances is the expectation. Among different metrics of similarity between the two arrays, Assuming that two overlapping pixel-arrays are sections of the same simultaneously observed earth region, high similarity between the two arrays of TOA radiances is the expectation. Among different metrics of similarity between the two arrays, we chose normalized Root Mean Squared Deviation:

\begin{equation}
RMSD = \sqrt{\frac{1}{N} \Sigma_{i=1}^{N}(TOARAD_{L9,i}-TOARAD_{L8,i})^2}
\end{equation}

Where N is the number of pixels in each array and TOARADL8 and TOARADL9 are the TOARAD for L8 TIRS and L9 TIRS, respectively. The RMSD is then normalized to the full range of the covered radiances in the corresponding scene pair. All the illustrated RMSD values displayed in diagrams throughout this work are normalized RMSDs. 
The parent dataset of 408 overlapping images was drawn from the dataset compiled by Choate et al and used for geometric and radiometric performance comparison of Landsat 8 and 9 OLI instruments \cite{cho22,gro22}. Choate et al. measured a 2.2m mis-registration between L8 and L9 coincident acquisitions between the OLI sensors which translates to less than one-tenth of a 30m multispectral pixel. Since alignments between the OLI and TIRS sensors are well calibrated, the mis-registration errors between the thermal images from L8 and L9 can be considered negligible \cite{cho22}. 

The Landsat instruments preferentially collect data over land surfaces as seen in Figure \ref{fig2}. The parent dataset was primarily selected to sample different land cover types for the VSWIR bands of OLI. Ideally, extended water scenes, such as the open ocean, would be used for comparison of the TIRS bands. However, the wider range and spatial variability of surface emissivity and temperatures within the parent dataset can yield useful insights into the performance of the TIRS bands. 
The selected scene pairs used in this study are a subset of the parent dataset. Figure \ref{fig2} shows both the parent dataset (blue open squares) and selected scene pairs (green squares).  The scene pairs have a wide range of overlap ($10-98\%$) and are spread across the globe. The process to determine the selected scene pair subset relies on is discussed in the following subsections.

\subsection{Scene pair selection}
A valid comparison between the two TIRS measured radiances is contingent on an accurate matchup between the geographical coordinates of the simultaneously observed scenes. In order to define a criterion for scenes that are qualified to be used in this study, the first step is to ensure that the comparison is limited to the areas that were simultaneously observed by the two sensors. Although Landsat L1TP images are $7k\times7k$ pixels in size, arrays of $1k\times 1k$ pixels from the overlapping sections of the image pairs were cropped so that the cutout size could remain the same even for the scene pairs with the lowest overlap ($10\%$). Visual vetting of a random set of cutouts ensured the reasonableness of the cropping process. In addition, the latitude and longitude coordinates of the cutout pairs should match. The pixel quality masks of each image (\texttt{LC0\_*\_QA\_PIXEL.TIF}) were also cropped in the same fashion and were later used to only select pixels that are identified as clear ($bitmask==21824$) in both cutouts of each scene pair.  The corresponding MTL files for each image contain the calibration coefficients to calculate TOA radiance and Brightness Temperature. The two $1k\times1k$ pixel arrays of radiance and temperature from each of the images in the pair, are then ready to be compared. Figure \ref{fig3} shows two examples of this cutout process. Top three panels of Figure \ref{fig3} are cloud free image cutouts (WRS2 path/row 30/40) of band 11 TOA brightness temperature of L8 TIRS (left), L9 TIRS (middle) and difference between the two (right). Similarly, lower panels of Figure 3 show the cutout (WRS2 119/38) for a scene with a large number of pixels classified as non-clear. A cursory visual inspection of the paired TIRS cutouts indicates good agreement in the geographic features observed and consistency in the non-clear masking for the two images. 

\subsection{Examined, Eligible, and selected scene pairs}

The parent dataset of 408 scene pairs went through the cropping process that involved defining a $1k \times 1k$ crop window in the overlapping area of each scene pair. This parent sample is also referred to as the examined scene pairs throughout this work. Despite the fact that both L8 and L9 platforms were coordinated to fly on the same WRS-2 (PATH/ROW), the assigned PATH of the two platforms may not be the same in every scene pair. Depending on how close the scene centers are to each other, the two scene’s PATH in a pair may be different by one (e.g.,133/40 and 132/40 are a pair of overlapping scenes). In conventional GeoTIF cropping routines, the crop window is defined in Universal Transverse Mercator (UTM) coordinates. Scene pairs with different PATHs can belong to different UTM zones, causing them to have mismatched UTM coordinates which results in cropping failure. It should be noted that using grid transformation algorithms that reproject the UTM coordinates of one image to those of the other image, is a common solution to resolve this cropping issue and salvage the data in studies in which omitting these image pairs critically reduces the size of the dataset. Caution has to be taken when using these transformed images for pixel-to-pixel comparison, as these algorithms include a resampling process that may affect the computed DN values.
 The image pairs that required this reprojection step to resolve their cropping issue were excluded from the dataset in this work. The eligible dataset is the remaining 301 scene pairs with successfully cropped cutouts. Finally, the selected scene pairs are a set of 157 scene pairs for which their calculated RMSDs of TOA radiances were lower than 5 percent.  To justify the selection of the $5\% RMSD$ threshold for image similarity, an error cumulative distribution analysis was conducted across a representative dataset of image pairs. This analysis demonstrated that an RMSD below $5\%$corresponded to the point at which over $95\%$ of visually similar image pairs clustered, providing a practical and statistically supported cutoff. We note that when comparing two measurements of the same target with two similar sensors such as the two TIRS in this study, measurements with RMSDs of $5-10\%$ are still considered similar. Relaxing the upper limit to $\rm RMSDs<10\%$ would increase the number of selected pairs to 233. This could be a viable option for data for similar studies with smaller parent dataset.

Throughout this work, only the pixels that were labeled clear in both L9 and L8 image cutouts were used to measure the RMSD of their TOA radiances.  No additional masks were applied to exclude the non-clear pixels besides the available QA pixel maps per L1TP image in the public USGS Landsat data archive. This $\rm RMSD < 0.05$ metric was chosen as an initial scene pair selection criterion with the possibility of refinement if there were scene pairs that exhibited sharp differences in pixel-to-pixel comparison metrics despite meeting this criterion.

\begin{figure}[!t]
\centering
\includegraphics[width=2.5in]{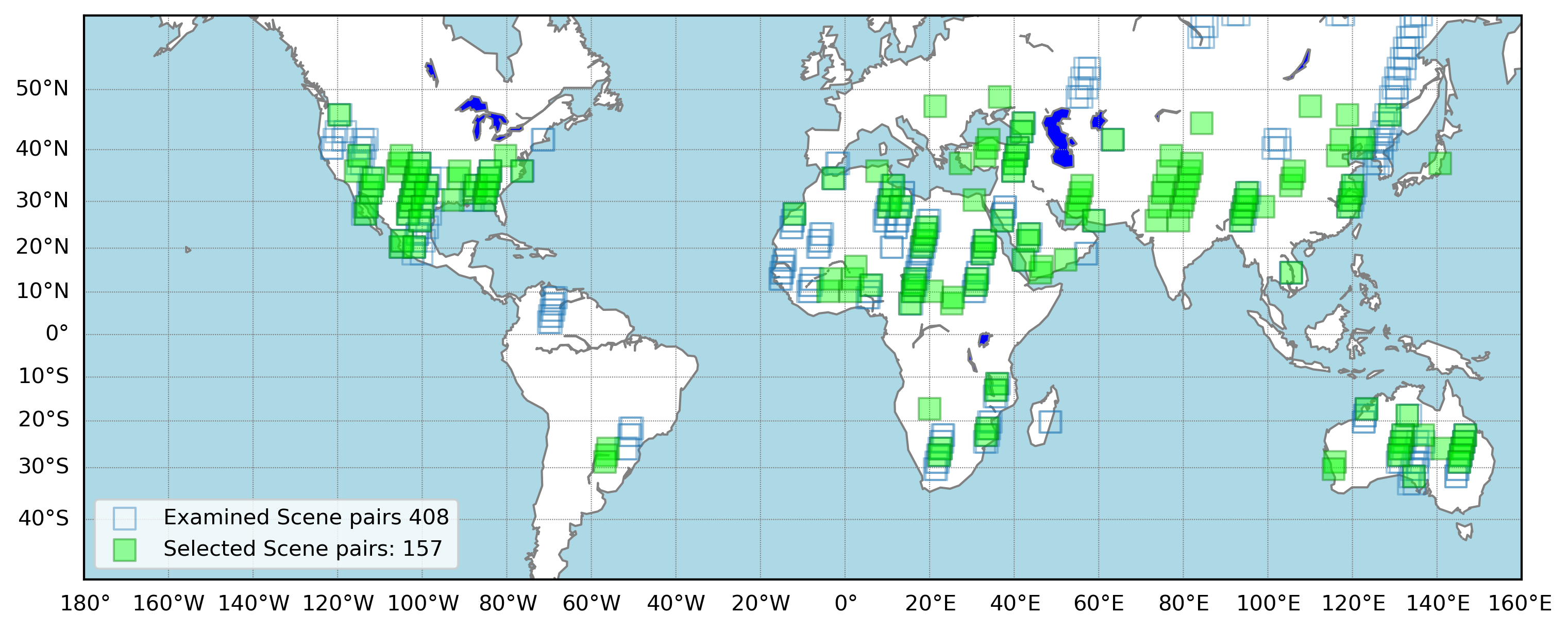}

\caption{Global coverage of the 408 scene pairs in the original sample used for this study (open blue squares), and the selected subset of those scene pairs that exhibited observing condition similarity for unbiased pixel to pixel comparison.
}
\label{fig2}
\end{figure}

\begin{figure}[!t]
\centering
\includegraphics[width=2.5in]{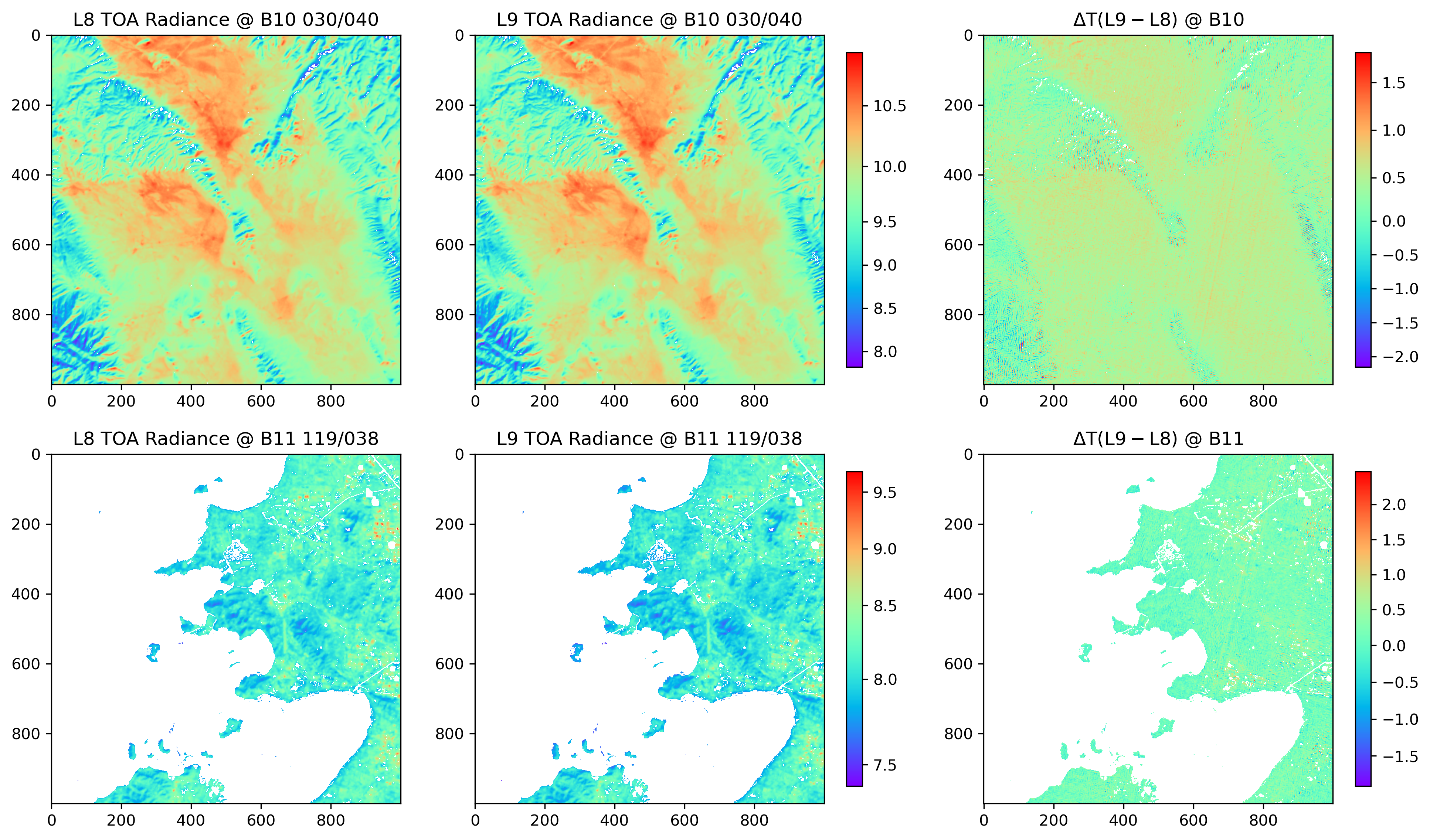}

\caption{Matched cutouts of TIRS and TIRS2 images that are precisely cropped to the same latitudes and longitudes in the coincidentally observed scenes with Path/Row of 30/40 and 119/38. The color-mapped cutouts show the TOA radiances at B10 (top) and B11(bottom). The right-most panels on the top and bottom rows are their pixel-to-pixel brightness temperature differences. Pixel quality maps have been applied to the images and all the pixels that are not categorized as clear are removed from both cutouts.}

\label{fig3}
\end{figure}

\begin{figure}[!t]
\centering
\includegraphics[width=2.5in]{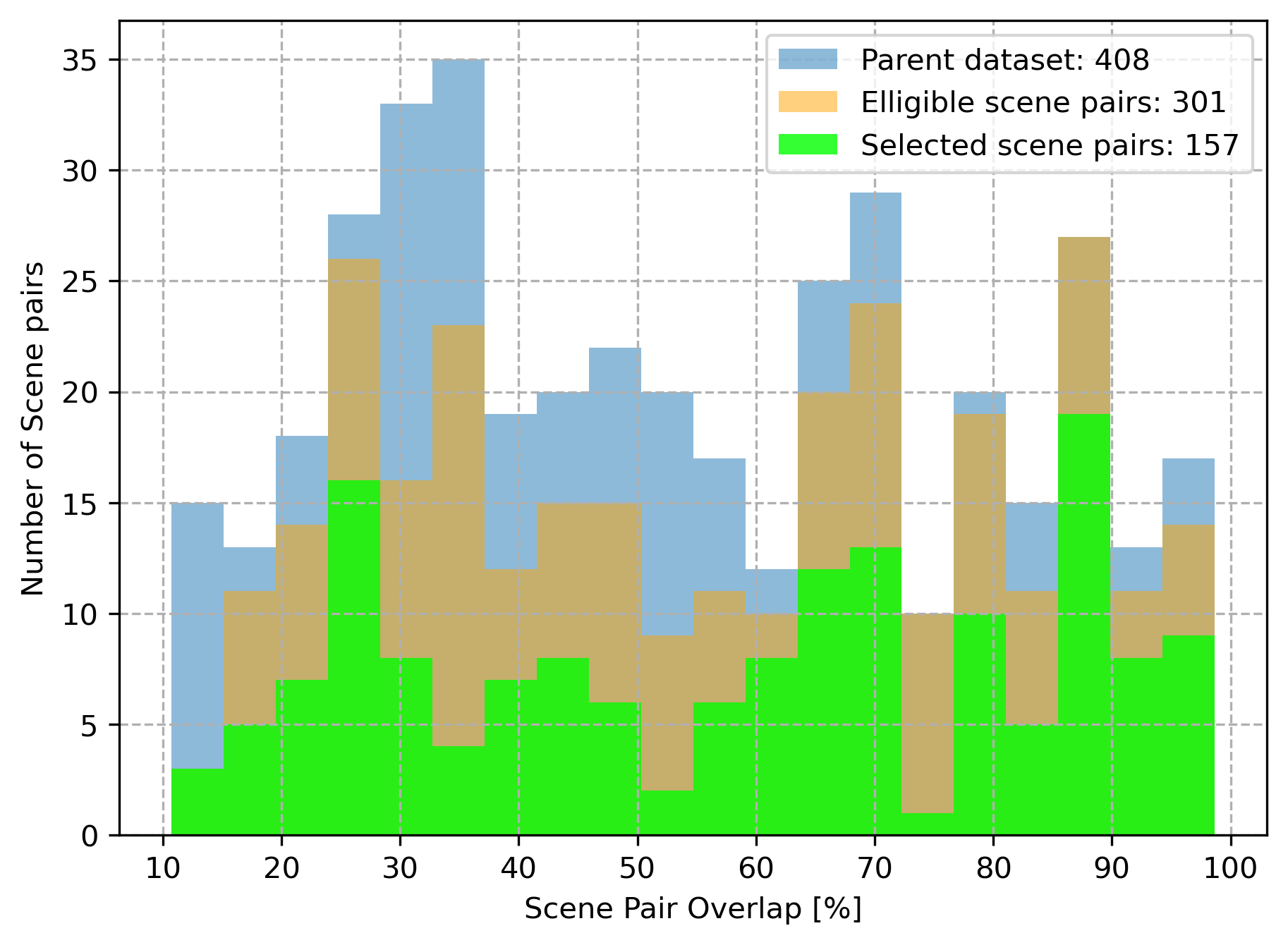}
\caption{Selected scene pairs (green) cover the full range of overlaps as the pairs in the parent sample of all scene pairs. This is one illustration of the notion that the selection criterion is not biased towards highly overlapping scene pairs. Eligible scene pairs are a subset of the parent sample for which a fit to their radiance and temperature distributions did not fail but implied dissimilarities between the two distributions.}
\label{fig4}
\end{figure}

\section{Results and Discussion}
As seen in Figure \ref{fig2}, the global spread of the selected scene pairs is primarily over continental areas. While additional overlapped areas exist throughout the multiple orbits during the underflight period, the parent set of 408 overlapping scenes identified by the Landsat team is of sufficient and manageable size for evaluation of the underflight analysis methodology. After applying the selection criteria described in the previous section, 157 scene pairs were identified for further characterization. That an ample number of scene pairs met the $\rm RMSD < 5\%$ metric, is an encouraging indicator that the pixel-to-pixel matchup process was correctly applied. One advantage of using both land and water scenes for thermal sensor comparisons is the wider range of possible scene radiance and brightness temperatures than if using water scenes only.
\begin{figure}[!t]
\centering
\includegraphics[width=3in]{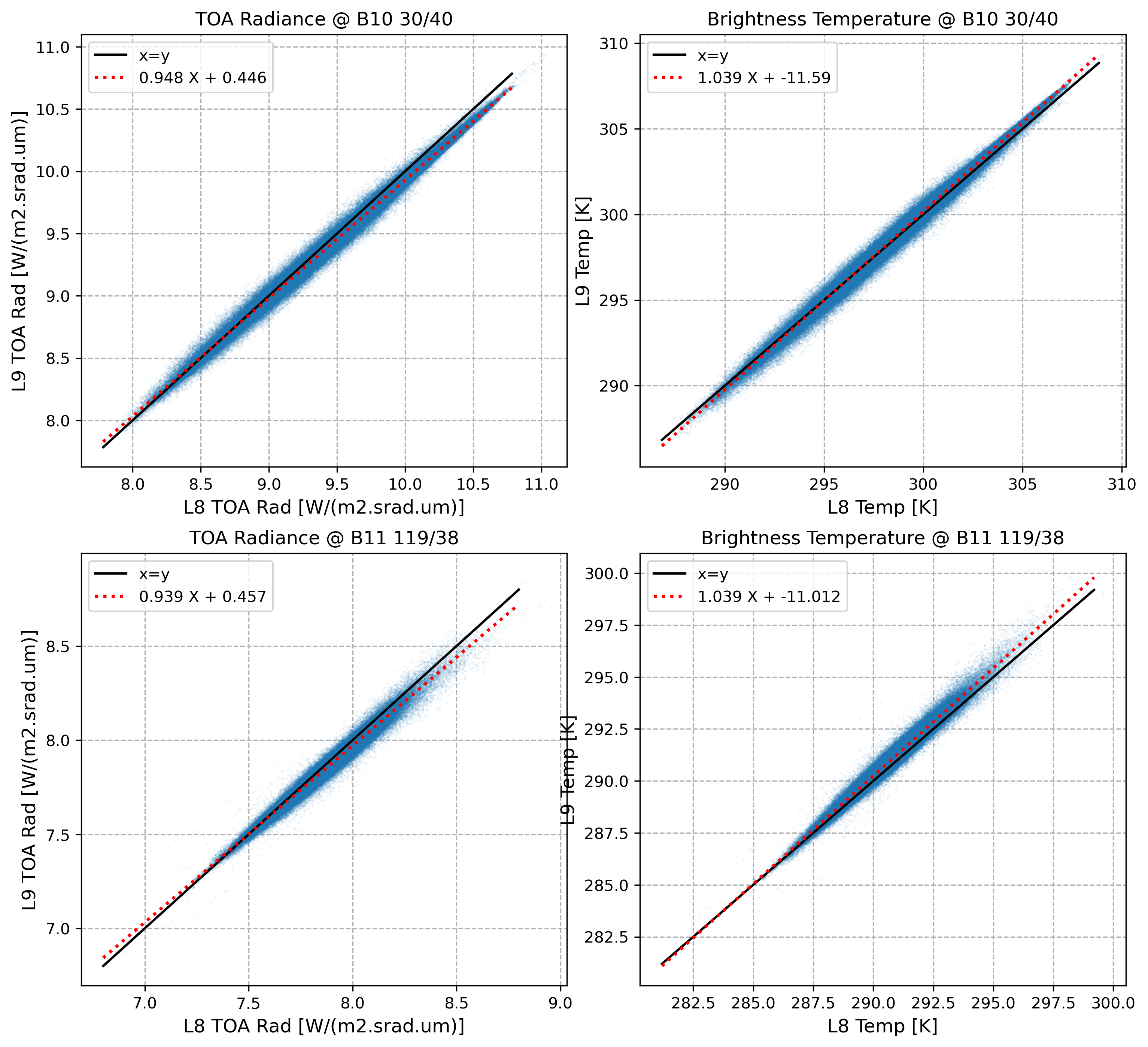}
\caption{Comparisons for the two example cutout image pairs shown in Figure \ref{fig3}. The observed good agreement with the equality line is not unexpected.}
\label{fig5}
\end{figure}

\begin{figure}[!t]
\centering
\includegraphics[width=3in]{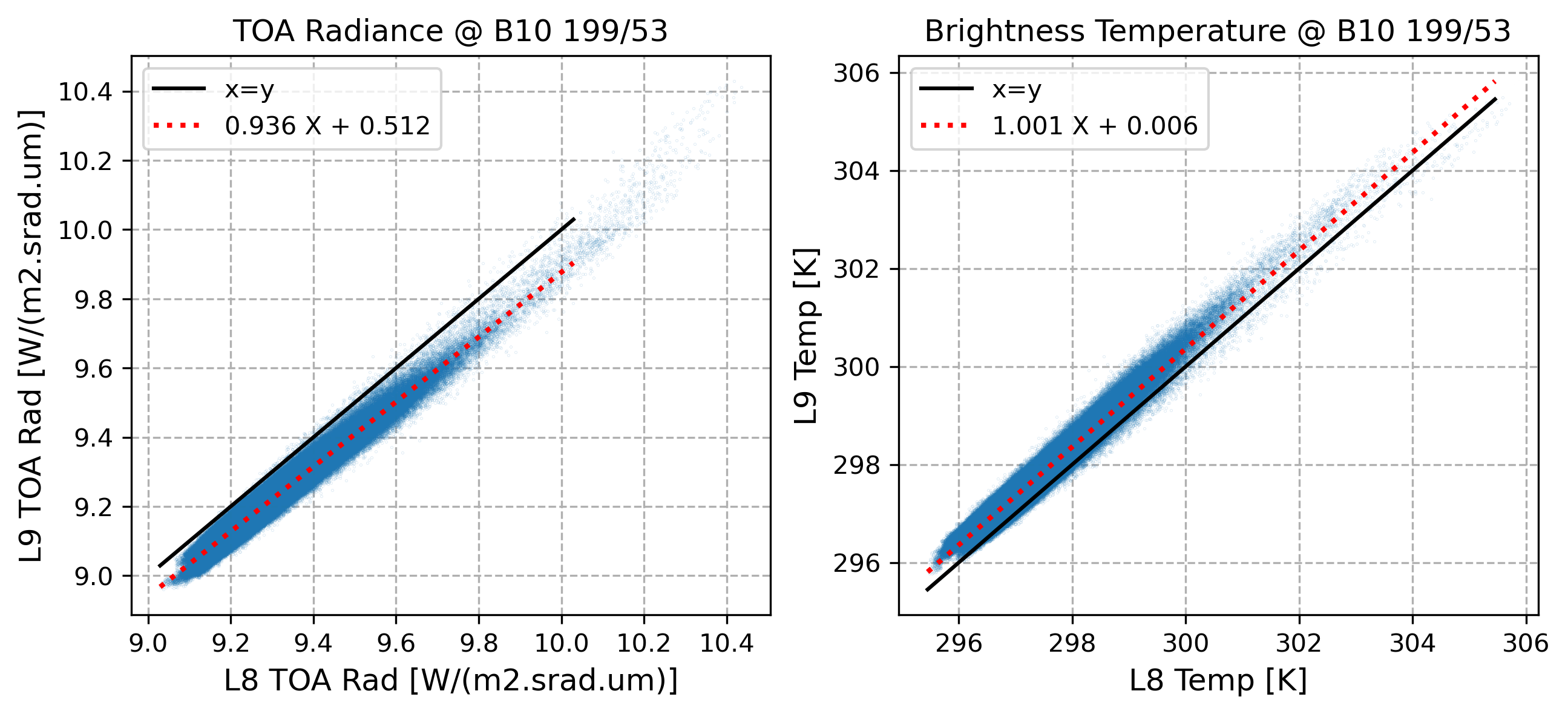}
\caption{An example of an eligible scene pair that displays an offset from the equality line, however the linear fit of the comparison is still good.}
\label{fig6}
\end{figure}

\begin{figure}[!t]
\centering
\includegraphics[width=3.5in]{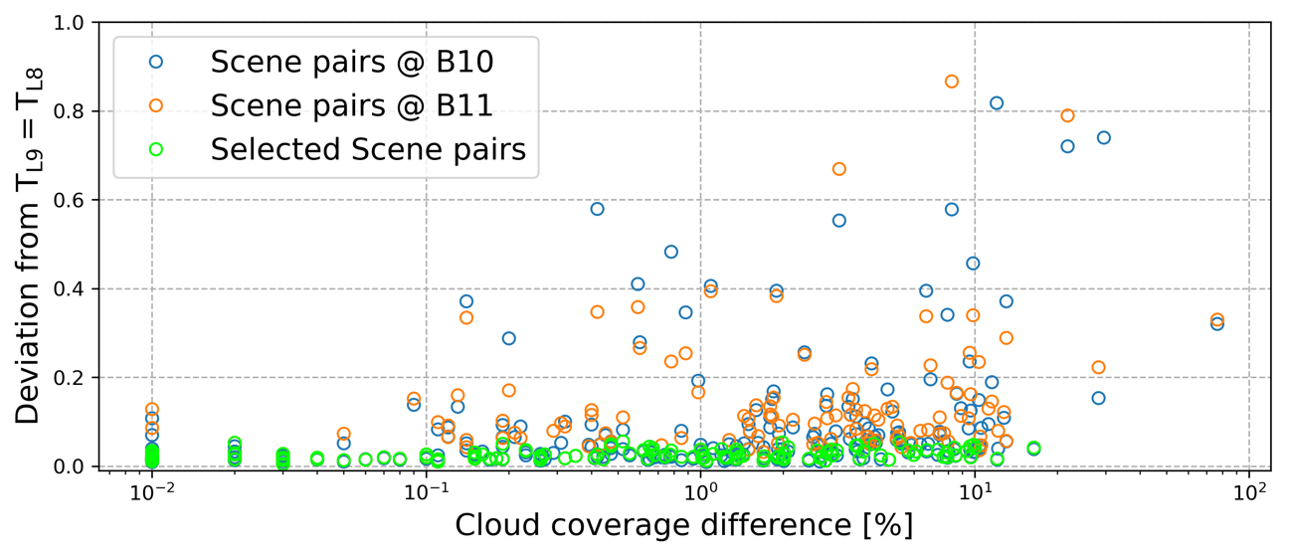}
\caption{An illustration of the robustness of $\rm RMSD<5\%$ criterion as a scene pair selection metric. The selected scene pairs (green) also have the lowest difference in the measured temperature by the two thermal sensors at both B10 and B11. Deviation from temperature equality is calculated as root mean square difference between the measured and expected (equal) value for temperature.}
\label{fig7}
\end{figure}

\begin{figure}[!t]
\centering
\includegraphics[width=3.5in]{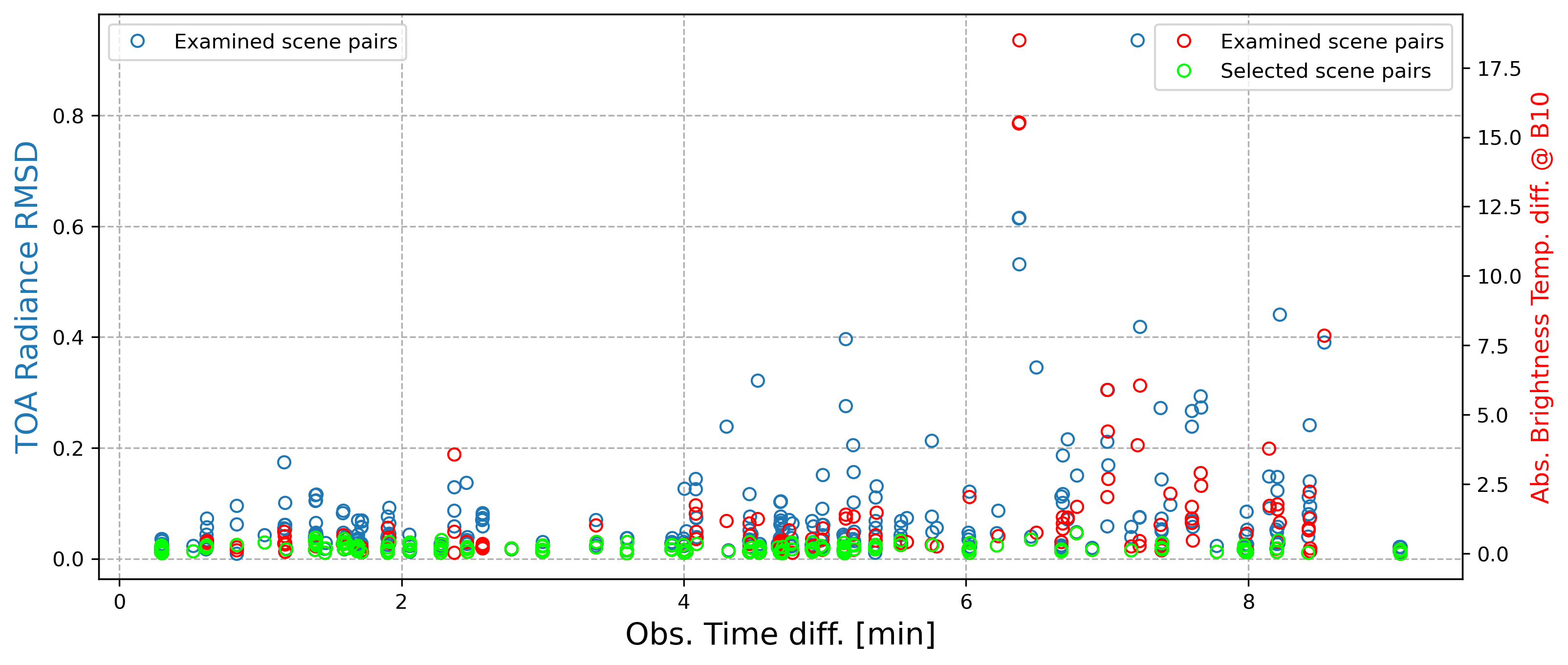}
\caption{Acquisition time differences of longer than 4 minutes appear to increase the likelihood of changed atmospheric conditions that impact the measured spectral radiance (e.g., fast moving clouds or optically invisible moisture patches moving into the field of view during the time gap between the two observations).}
\label{fig8}
\end{figure}

\begin{figure}[!t]
\centering
\includegraphics[width=3.5in]{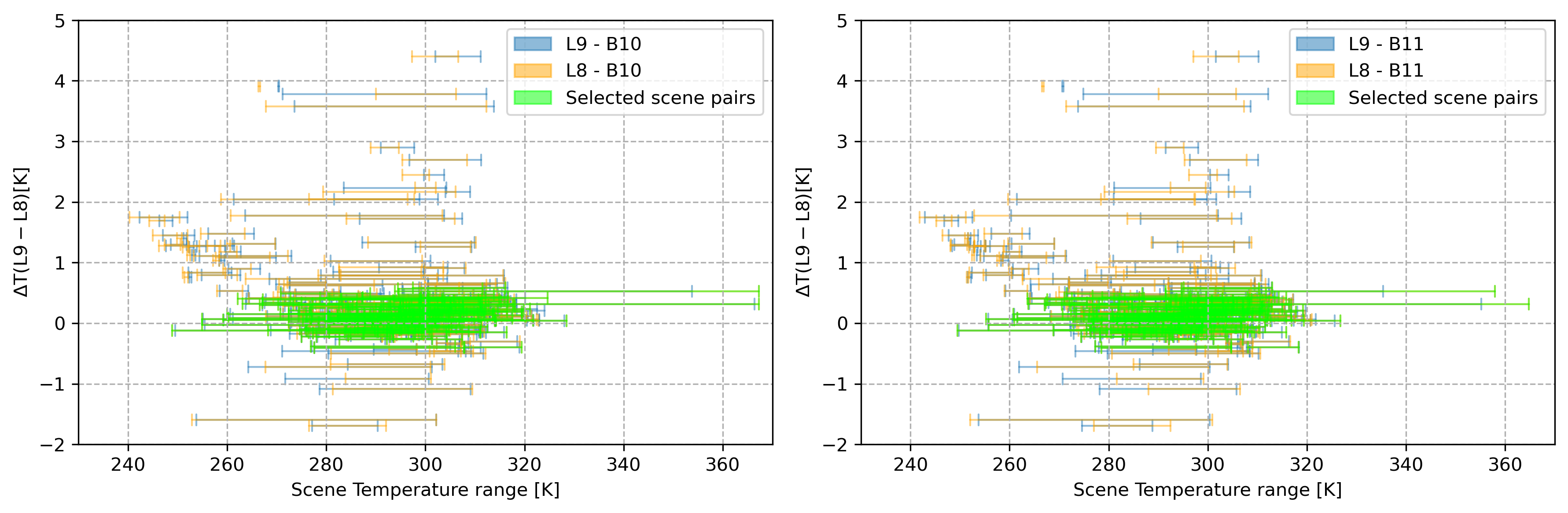}
\caption{Temperature range of the scene pairs in the parent sample versus the selected scene pairs (green) at B10 and B11.  Selected scene pairs span over as wide of a scene temperature ranges as those of the full sample, and yet the selected scene pairs remain in the near zero temperature difference. Note a population of scene pairs with small temperature range at $< 270K$, with $T_{L9} > T_{L8}$. This hints at an overcorrection in stray light mitigation algorithm at the cold end of the measured TOA brightness temperatures.}
\label{fig9}
\end{figure}

\begin{figure}[!t]
\centering
\includegraphics[width=3.5in]{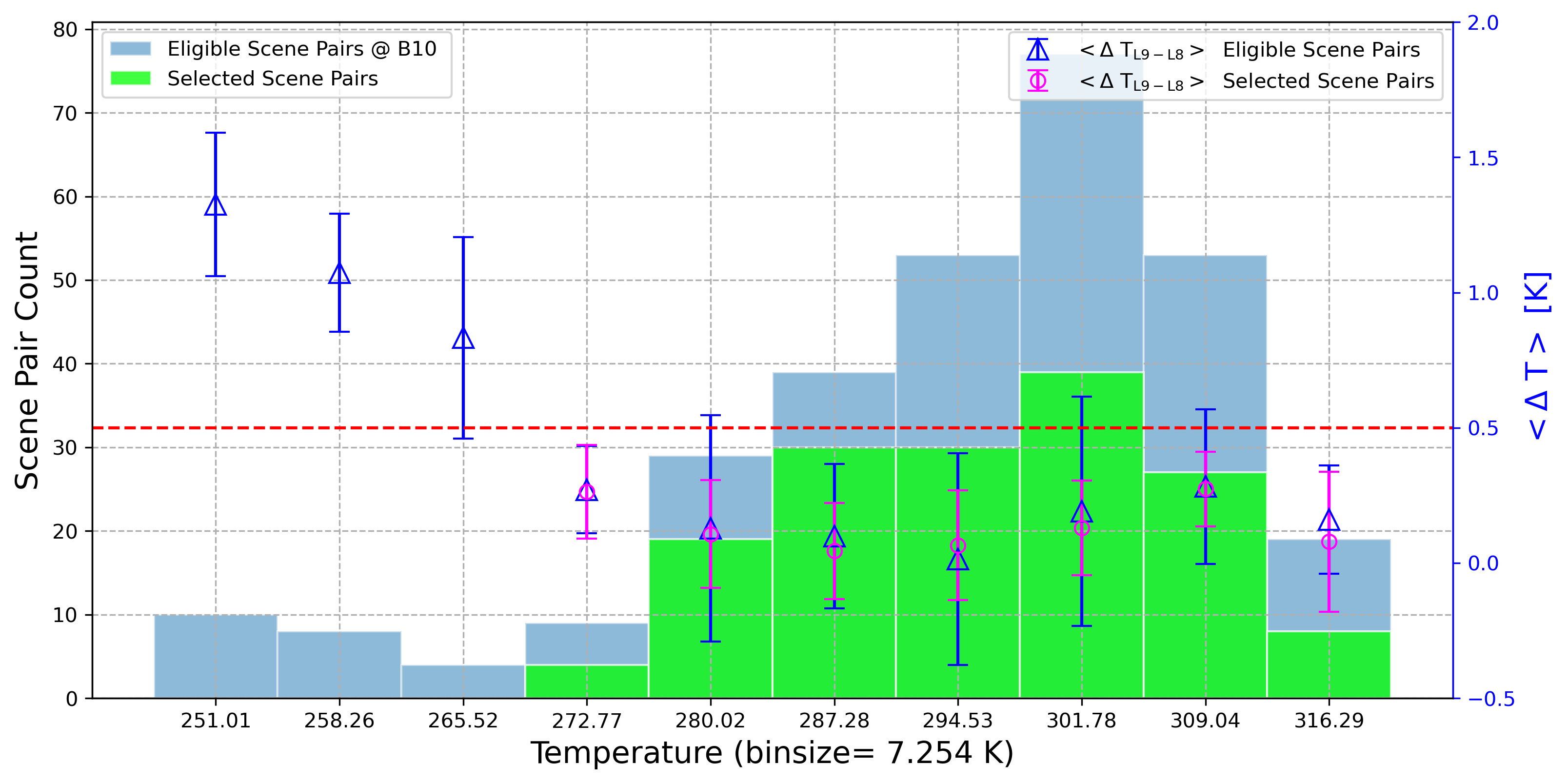}
\caption{The averaged pixel-to-pixel temperature difference between two measurements by TIRS and TIRS-2 exceeds 1K at $< 260K$. Selected scenes are well within the system requirement margin of 0.5K. }
\label{fig10}
\end{figure}

\begin{figure}[!t]
\centering
\includegraphics[width=3.5in]{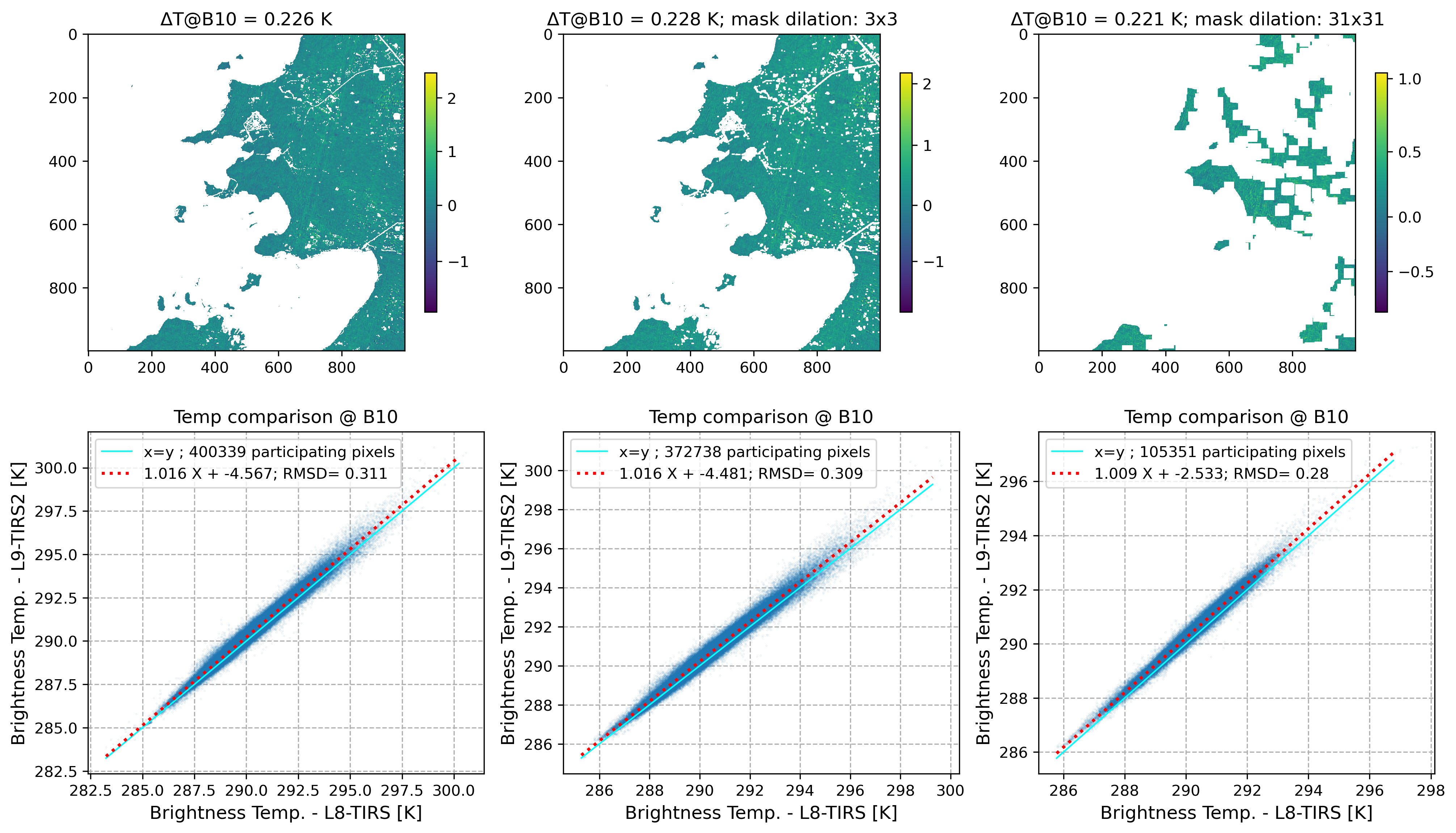}
\caption{An illustration of pixel quality mask dilation effect on temperature comparison.}
\label{fig11}
\end{figure}

\begin{figure}[!t]
\centering
\includegraphics[width=3.5in]{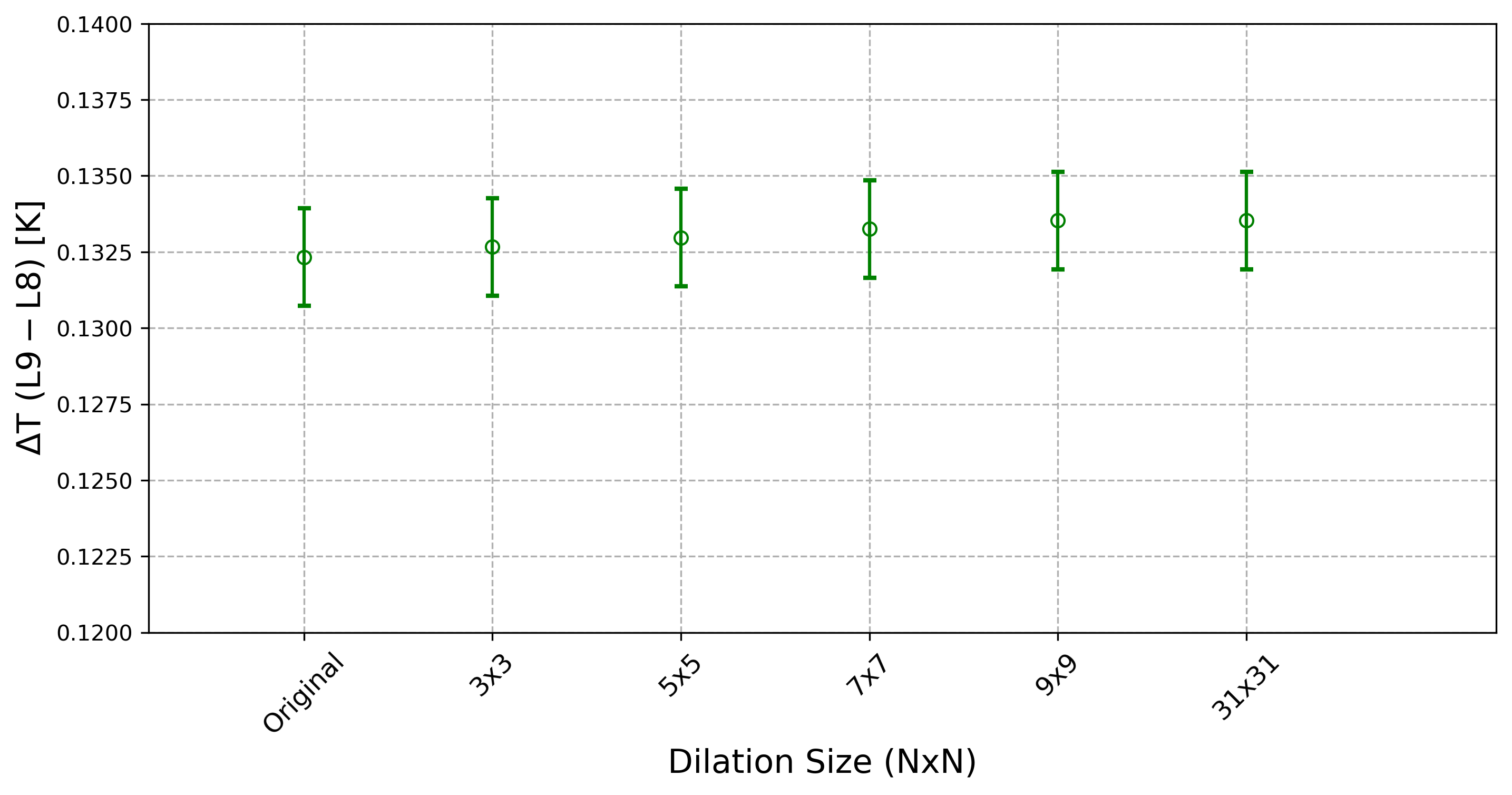}
\caption{Average$\rm \Delta T$ versus the pixel dilation size. There is minimal impact on the derived$\rm \Delta T$ when averaging a large number of pixels}
\label{fig12}
\end{figure}

\begin{table}[!t]
\renewcommand{\arraystretch}{1.3}
\caption{Average and median of differences in TOA radiance ($DR=TORAD_{L9}-TOARAD_{L8}$) and Brightness Temperature ($\Delta T=T_{L9}-T_{L8}$) for the sample of 157 ``selected’’ and 301 ``eligible’’ scene pairs}
\label{table1}
\centering
\begin{tabular}{|l|c|c|}
\hline
Dataset & $\rm \Delta R ~ \rm mean (median)$ & $\rm \Delta T ~ \rm  mean (median)$  \\
& $[Watt/m^2.srad.\mu m]$ & $[K]$ \\

\hline
Selected Pairs at B10 & 0.014 (0.043) & 0.123 (0.134)\\
 at B11 & -0.042 (-0.012) & 0.066(0.057)\\

\hline
Eligible Pairs at B10 & 0.241 (0.057) & 0.0965 (0.203)\\
at B11& 0.170 (-0.015)& -0.0093 (0.086)\\
\hline
\end{tabular}
\end{table}


One concern with the pair selection process is a possible selection bias in the selected set of scene pairs versus the parent dataset. As an example, Figure \ref{fig4} shows a histogram of the parent (blue), and selected image pairs (green) based on the percentage overlap of the paired images. It is evident that the selected pair distribution is consistent with that of the parent sample across the whole range of overlap percentage. A similar result was seen for histograms of cloud coverage percentage detected in the image pairs. This gives confidence that a meaningful result can be obtained even for image pairs with minimal overlap $(\sim10\%)$. 

For each selected pair, a comparison of the pixel-to-pixel radiance and brightness temperature between the two TIRS instruments was generated. Figure \ref{fig5} shows these comparisons for the two example cutout pairs shown in Figure \ref{fig3}. The observed good agreement with the equality line is not unexpected.

The majority of the selected scene pairs were seen to have their per-pixel values symmetrically distributed around the equality line with slopes and intercepts closer to one and zero respectively. Although the comparisons are drawn from clear pixels only, losing a large portion of total pixels to various pixel quality flags (e.g. bottom panels in Figure \ref{fig3}) does not disqualify a scene pair from use. Scene pairs with higher number of clear pixels garner more statistically confident comparisons, but this does not preclude that scene pairs with less available matched pixels do not provide valid comparison results.

Figure \ref{fig6} shows an example of an eligible scene pair that displays an offset or bias from the equality line. Although the linear fit of the comparison is good, and the image pair barely misses the $\rm RMSD <5\%$ criteria (with $\rm RMSD\sim 0.055$), the potential calibration offset causes the average pixel-to-pixel temperature difference between the scene pairs to soar to 0.36K. This$\rm \Delta T$ is below the system requirement threshold (0.5K) but more than three times higher than the average temperature difference of the selected scene pairs ($\Delta T \sim 0.1 K$). Such scene pairs do not lose their eligibility for inclusion in the dataset for having an offset to the distribution of per-pixel radiances or temperatures with respect to the equality line. These scenes are eligible for relative calibration with respect to the confidently calibrated sensors especially in smaller underflight datasets with fewer eligible scene pairs in nominal observing conditions.  

Quantifying the degree of similarity for pixel arrays of calibrated TOA radiances over the same region provides a quick and easy-to-measure metric that is applicable even when used for less similar sensors than the near-identical sensors in this study. It should be emphasized that the fact that the TIRS sensors aboard L8 and L9 are near-identical in their technology, design, detection and operation is a key factor in the sanity of this assessment and the eligibility of this underflight dataset to formalize the similarity criteria provided here. 
Figure \ref{fig7} shows the RMSD for per-pixel TOA radiances of TIRS for B10 for the parent dataset (blue) and that of the selected pairs (green) are shown versus the cloud coverage difference between the image pairs. The cloud coverages are taken from the stored values in CLOUD$\_$COVER keyword in MTL files of the associated L1TP images (similar pattern is observed for B11). As the parent dataset was originally selected for analysis of the Landsat OLI comparisons, the overall low cloud coverage $(<10\%)$ in the pairs is expected.
This figure illustrates that scene selection simply based on RMSD of TOA radiances, is expectedly, equivalent selection of most cloud-free images. The presence of scene pairs with $\rm RMSD>0.05$ in the parent dataset is indicative of potential changes in the scene observed by the two sensors. While the observations are nearly simultaneous there will exist time differences in the acquisitions, which allows for the possibility of a change in cloud presence, surface temperature, or atmospheric conditions between the two.

The near-coincident observations during the full extent of the underflight maneuver vary in their relative acquisition time gap. This time gap remains under 10 minutes for all the scene pairs in the parent dataset used in this work. Figure \ref{fig8} shows the difference in average brightness temperature ($\Delta T$) between the image cutouts in a pair, for eligible scene pairs (blue) and selected scene pairs (green) for B10 (similar pattern is observed for B11). There appears to be a noticeable increase in the measured$\rm \Delta T$ for the paired L8 and L9 acquisitions that are $>4$ minutes apart.  In general, the shorter the observation time difference, the more likely to have a low RMSD for the image pairs. For longer observation time differences, greater than 4 minutes, there is a greater likelihood of a change in atmospheric and surface conditions or cloud coverage between the image pairs. This is evidenced by the greater scatter on the right side of the plot.  As is shown in Figure \ref{fig8}, the selected scene pairs dataset includes scene pairs with $>4$ minutes acquisition time difference but still exhibit $\rm RMSD<5\%$. This highlights that longer acquisition time difference is not an immediately disqualifying factor for a scene pair but a facilitator for other environmental causes for dissimilarity to impact the sensor\'s measurement. 
Calibration accuracy budgets are generally temperature-dependent. Any sensor intercomparison benefits from including data across the full width of the sensor’s dynamic range. Figure \ref{fig9} displays the scene temperature ranges observed for both the eligible and selected datasets. As expected, the selected image pairs cluster around a scene temperature difference of zero Kelvin, with an average$\rm \Delta T$ of 0.1K.

The scene temperature ranges from $\sim 250-360$ K for the selected dataset. Anomalous data points such as excessively hot and cold pixels or cloud-affected scenes do not strongly impact the comparison results because of the relatively small number of such data pixels. Additionally, the Landsat 9 orbit during underflight drifted from cases where the Landsat 9 data were acquired prior to Landsat 8 to those after Landsat 8. Thus, a large fraction of effects that are temporally dependent lead to both positive and negative differences.

The selected image pairs in this dataset were binned into ranges of $<10K$, shown in Figure \ref{fig10}. The particular bin size (7.25K) was optimally chosen to ensure that the full extent of the covered temperatures in both scene pair images still kept both pair members in the same bin of temperature. The blue histogram shows that most of scene temperatures in the sample fall within the typical range of $280-305 K$, but that there are some lower scene temperatures included. The overplotted data points are the averaged temperature differences of the scene pairs that fall within the extent of each bin. It is of interest to note the larger temperature differences ($>1K$) at the colder scene temperatures. 
Landsat 8 TIRS has a well-documented stray light issue \cite{mon14,ger15,mon22}. The correction algorithm applied during data processing is optimized for the typical scene temperature range $\sim 280-310K$ so the observed temperature dependence may indicate an over-correction at low scene temperatures. The cluster of points top-left section of Figure \ref{fig9} show those colder scene pairs in the parent dataset with$\rm \Delta T \sim 1-1.5 K$.

Since it is impossible to operationally quantify the out-of-field signal for any given scene, the stray light correction is less effective for colder scenes.  As this work focuses on comparing pixels that are identified as clear in both image cutouts, the selected scene pairs remained unaffected by the level of cloud coverage (Figure \ref{fig7}). But one takeaway from the illustrated impact of the observed time gap between pair acquisitions in Figure \ref{fig8} is the possibility of the cloud coverage or atmospheric properties changing within the sensors’ field of view between the acquisitions. This motivated a similarity improvement assessment when dilated masks of various sizes are applied. Dilated masks remove the pixels with non-clear neighbors in $3\times3$ (a window representing a similar size to the at-sensor 100-m pixel size of thermal images before the up-sampling process in the pipeline for L1TP images), $5\times5$, $7\times7$ (representing one 100-m pixel size away from a poor-quality pixel in all directions), $9\times9$ and $31 \times 31$ windows.  Figure \ref{fig11} shows an example of the effect of applying dilated masks on the temperature similarity level between the paired images shown in the bottom panels of Figure \ref{fig3}. The top three panels in Figure \ref{fig11} show the same scene image for no dilation (left), $3\times3$ dilation (middle), and $7\times7$ dilation (right). The impact on the image is evident in the decreasing number of available pixels. The bottom three panels are the companion pixel-to-pixel comparisons for the aforementioned dilation cases and illustrate the relatively minimal effect of this process.

Despite the notion that dilation removes stranded clusters of pixels and could positively affect the per-pixel temperature similarity (by removing pixels with higher uncertainty in temperature measurement due to their proximity to cloudy, icy or snowy pixels), the potentially positive effect can also bias the distribution of pixel values toward the warmer end and increase the average temperature of the remaining pixels. Figure \ref{fig11} displays the average$\rm \Delta T$ versus the pixel dilation size. There is minimal impact on the derived$\rm \Delta T$ when averaging a large number of pixels, in this case the$\rm \Delta T$ difference between a single pixel and the $31\times31$-pixel average is 0.0012 K. 
Table \ref{table1} lists the average and median of the pixel-to-pixel TOA radiance and Brightness Temperature differences ($\Delta R$ and$\rm \Delta T$) for B10 and B11.

One motivation for this study is that there is an increased interest for future missions to use underflight maneuvers operationally as one of the means of harmonizing similar measurements from multiple sensors. Coordination by mission teams in orbital planning can result in underflights at launch but also periodically once in final orbit. This idea is also relevant to the proliferating numbers of SmallSats which typically don’t include extensive on-board calibration capabilities. For any given sensor in orbit there will be underflights-of-opportunity with a growing number of sensors. 
There are surmountable challenges in predicting these underflights-of-opportunity for a large number of sensors. The ability to harmonize the measurements of multiple sensors to a common scale would increase the utility of the collected datasets to the world-wide science community. 
Analysis of the dataset from Landsat 9’s underflight of Landsat 8 provided a useful test case of a method for intercomparing the two calibrated thermal sensors. One advantage of using two near-identical TIRS instruments is the elimination of complications of differing instrument specifications in the comparison. Having numerous orbits of overlapping data, provided a set of diverse terrain types and regions with varying spatial and thermal homogeneity and overall range of the covered temperatures. Combining underflight datasets with conventional approaches using instrumented sites, streamlines the calibration validation process, adds statistical value by adding globally-distributed, non-instrumented sites, making the process more time-efficient and less resource-intensive. Periodic underflight maneuvers that pair an orbiting sensor with an already verified calibration with a newly launched platform, provides a significant number of scenes at a wider range of surface temperatures for a comprehensive calibration comparison during the commissioning period of the newly commissioned sensor.  
The near-coincidental observations of the underflight data eliminate the time-variable component and the associated uncertainties that come with it. Therefore, a detected anomaly in the calibrated data over the same area can be more directly connected to instrumental performance defects rather than the observed source. Expanding the introduced selection criteria in this work to other wavelength ranges could provide a spectral understanding of its applicability across different remote sensing channels. Developing automated tools and algorithms to implement the wide-net filter approach can further enhance practicality, facilitating rapid identification and qualification of calibration sites and making the process more accessible and scalable.
The criteria of accurate geographical coordination, reasonably deferential pixel mask with clear, ice and cloud bitmasks, and exclusion of features with rapid thermal variability enable a broad range of scenes to qualify for calibration validation and comparison, regardless of their terrain characteristics and land-water fraction.

\section{Conclusion}
This study is an attempt to assess the feasibility and the extent to which the “full orbit data” from the underflight maneuvers can be used for harmonization of participating sensors’ calibrations. Not only scenes with a wide range of overlap ($10-98\%$) can be used in this approach, partial cloud coverage and presence of ice/snow in the scene would not immediately disqualify a scene pair for calibration assessment.  By using reasonably inclusive bitmasks and limiting the pixel-to-pixel comparison to only clear pixels with clear immediate surroundings, partially clear scenes can also yield impressive temperature agreement, below the system requirement. The examined scene pairs for this study are spread over a wide range of geographical locations, terrain types, water-land percentage, scene temperatures (250 to 320K) some of which include over 50K temperature difference in one scene. We observed temperature agreements of within 0.123 K and 0.066 K for the 10.9 $\mu m$ and 12.0 $\mu m$ bands. 

 This work highlights that (i) A diverse dataset in all these aspects, strengthens the argument for using the full orbit data over areas with wider surface temperatures than the conventionally utilized, instrumented Cal/Val sites. At the very least, full orbit data can augment Cal/Val datasets and flag the potential cold/hot biases; (ii) Similar observing and atmospheric condition for the near-coincidental observations are key factors in fair comparison of the underflying sensors; (iii) Significant temperature differences ($>0.5K$) between the scene pairs have been always associated with acquisitions that are more than four minutes apart; (iv) Finding useful and qualified scene pairs for calibration comparison does not necessarily have to be limited to the scenes with perfect pixel-to-pixel radiance match (slope and intercept of 1 and 0 for the linear fit to their pixel-to-pixel comparison plot). Scene pairs with pixel-to-pixel TOA radiance distributions that converge tightly around different lines (rather than x=y line) could be used for relative calibration (where one sensor becomes the basis for calibrating the other).


%



\section*{Acknowledgment}
SEK is grateful to Landsat 8 and 9 calibration team for sharing a subset of their compiled set of scene pairs that became the parent sample for the scene pair selection scheme in this work. We thank Matthew Montanaro for sharing helpful insights on Landsat 8 calibration. 


\ifCLASSOPTIONcaptionsoff
  \newpage
\fi





\bibliographystyle{IEEEtran}
\bibliography{bibtex/bib/mybib} 

\end{document}